\documentclass[amsfonts, prd, twocolumn, nofotinbib, showpacs]{revtex4}
\usepackage{amssymb}
\usepackage{amsmath}
\usepackage{amsfonts}
\usepackage{graphicx, float}
\usepackage{graphicx, epsfig}
\usepackage{color}
\usepackage{enumerate}

\newcommand{\beq}{\begin{equation}}
\newcommand{\eeq}{\end{equation}}
\newcommand{\bea}{\begin{eqnarray}}
\newcommand{\eea}{\end{eqnarray}}

\newcommand{\Mbh}{M_{\rm BH}}
\newcommand{\SM}{M_\odot}
\newcommand{\Mpl}{M_{\rm Pl}}
\newcommand{\Ls}{L_*}

\newcommand{\ellp}{\ell_{\rm Pl}}

\begin{document}
\title{Extended Uncertainty Principle Black Holes}

\author{J. R. Mureika\footnote{E-mail: jmureika@lmu.edu}}
\affiliation{Department of Physics, Loyola Marymount University, 1 LMU Drive, Los Angeles, CA, USA~~90045}

 %%%%%%%%%%%%%%%%%%%%%%%%%%%%%%%%%%%%%%%%%%%%%%%%%%%%%%%

\begin{abstract}
An Extended Uncertainty Principle inspired Schwarzschild metric that allows for large scale modifications to gravitation is presented.  At a new fundamental length scale $\Ls$, the usual black hole characteristics (horizon radius, ISCO, and photosphere) deviate from their general relativistic counterparts by an additional term proportional to $\frac{G^3M^3}{\Ls^2}$ for $\hbar=c=1$.   If the scale is $\Ls\sim10^{13}$m, EUP modifications become relevant for black holes of mass $M \geq 10^{6}~M_\odot$.  This would affect the characteristics of most known supermassive black holes, and thus presents a unique set of experimental signatures that could be tested by the Event Horizon Telescope and similar future collaborations.  The Newtonian potential is similarly modified, and it is shown that for values of $\Ls$ in the range considered, the effect will emerge at about 1~kpc from the galactic center, coincident with the flattening of the Milky Way's rotation curve.  This suggests that the EUP could contribute to dark matter effects.  \end{abstract}

%%%%%%%%%%%%%%%%%%%%%%%%%%%%%%%%%%%%%%%%%%%%%%%%%%

\maketitle

\section{Introduction}
 Black hole physics is entering a golden age.  A century after Einstein proposed his celebrated theory of General Relativity, two key experiments are providing new data that promise to revolutionize the field: the Laser Interferrometer Gravitational Wave Observatory (LIGO) \cite{ligo}, and the Event Horizon Telescope (EHT) \cite{eht}.  LIGO has already ushered in the era gravitational wave astronomy, allowing us for the first time to probe the Universe beyond the electromagnetic spectrum.  Through this window, we will be able to test black hole mergers, coalescence, and ringdowns with high precision.  Conversely, the EHT is expected to reveal such novel characteristics as ``black hole shadows'' -- {\it i.e.} regions surrounding black holes from which no light escapes -- which will provide unprecedented tests of gravitational physics at the event horizon.

Quantum gravity effects are primarily important for the physics of microscopic black holes, including their evaporation profile and singularity removal.   There is growing consensus, however, that quantum characteristics of the underlying gravitational theory might also be relevant to macroscopic, large scale phenomena.  This is particularly important for the near horizon physics of astrophysical or supermassive black holes, including metric fluctuations \cite{giddings1,giddings2,giddings3,giddings4}, exotic compact objects like boson or Planck stars \cite{bosonstar,planckstar}, and quantum structures that exist just outside the horizon \cite{fuzz1,fuzz2,fuzz3},  It has been shown that such structures may even produce gravitational wave echoes in the ringdown phase of a compact binary merger \cite{Cardoso:2017cqb}.  In fact, it has been reported that such echoes have been observed with high statistical significance in the recent LIGO data \cite{niayesh1,niayesh2,niayesh3}, though these findings have been highly scrutinized \cite{moffat2}.

These ideas suggest that black holes are fundamentally quantum objects, regardless of their size.  Although the horizon is a prediction of General Relativity (GR), the physics governing it and the associated thermodynamics arise from quantum principles.  This is perhaps most apparent from the black hole's entropy,  which is given by the well-known area-entropy law, $S \sim \frac{A}{4}$, and not the expected volumetric scaling for a classical object.

The horizon itself may arise from a quantum origin.  For example, the corpuscular framework \cite{DvaliGomez} describes a black hole as Bose-Einstein condensate (BEC) of $N$ soft gravitons weakly confined to a region of size $R$, with total energy $E_{\rm tot}\sim M_{\rm BH} \sim N\frac{\hbar}{R}$.  The number of gravitons can be expressed as $N \sim \frac{R^2}{\ellp^2}$, and noting that $\hbar = m_P \ell_P$, the spatial extent of the BEC black hole is $R \sim \frac{\ell_P}{m_P} \Mbh \sim GM_{\rm BH}$, which is the expected relation. 

Of paramount interest to the EHT data is the morphology of the photosphere, more popularly known as the black hole shadow.  It has been shown that this feature can have a highly model-dependent shape (see \cite{shadow1,shadow3,shadow4,shadow5,shadow6,moffat,shadow8,shadow9,shadow10} and references therein).  The most striking prediction from a modified scalar vector tensor gravitational theory is that the shadow could be up to ten times the size predicted by General relativity \cite{moffat}.

Given the imminent release of EHT data, it is of timely interest to consider what other quantum effects might have an influence at extremely large scales.  In the following paper, the Extended Uncertainty Principle (EUP) will be considered as such a candidate mechanism.  An effective EUP-corrected Schwarzschild metric is constructed, and the associated black hole characteristics are derived.   It is shown that if the EUP contributions become important on scales of dark matter effects, there could be measurable deviations in the horizon radius, matter orbits and innermost stable circular orbit (ISCO), as well as the size of the photosphere of supermassive black holes (SMBHs) with masses $M \geq 10^6 M_\odot$.

\section{An Extended Uncertainty Principle Inspired Metric}
The Generalized Uncertainty Principle (GUP) \cite{mannkempf,Adler_1} is a quantum gravitational correction to the Heisenberg relation,
\beq
\Delta x \Delta p \geq 1 + \beta \ellp^2 \Delta p^2 
\label{gup1}
\eeq
where $\beta$ is a constant of order unity and dimensionless units $\hbar=c=1$ are assumed.  This is a common feature of disparate quantum gravity theories, including 
string theory~\cite{veneziano_1, veneziano_2, veneziano_3, veneziano_4, veneziano_5, veneziano_6},
loop quantum gravity~\cite{ashtekar_1,ashtekar_2},  
gravity ultraviolet self-completeness \cite{nicolini},
non-commutative quantum mechanics \cite{majid}, 
and generic minimum length models \cite{maggiore_1,maggiore_2,maggiore_3}.
The literature is rich with studies of GUP black holes and a full summary would be exhaustive, but the interested reader is referred to \cite{,Adler_2,Adler_3,gup1,gup2,gup3,mijmpn,gupextrad,bcjmpn} and references therein.  Such GUP effects are strictly relegated to the domain of Planck-scale black holes and their thermodynamics, as well as the Unruh effect \cite{gup4}.

The Extended Uncertainty Principle (EUP), on the other hand, introduces a position-uncertainty correction to the relation of the form \cite{mannkempf,bambi}
\beq
\Delta x \Delta p \geq 1 +\alpha \frac{\Delta x^2}{\Ls^2}
\label{eup1}
\eeq
Here, $L_*$ is some new, large fundamental distance scale, with $\alpha$ taken to be of order unity.  

EUP corrections have only recently been taken seriously, since the need for large-scale corrections to gravitational physics was not previously believed to be necessary \cite{bambi}.  This, of course, is no longer the case, so the EUP provides an avenue to introduce quantum effects over macroscopic distances.  As such, the EUP can modify the thermodynamics of FRW and (anti-)de Sitter spaces \cite{eup1,eup2}.  

In a previous paper \cite{bcjmpn}, the authors constructed a GUP-inspired metric of the form
\beq
ds^2 = F(r)~dt^2 - \frac{dr^2}{F(r)}-r^2 d^2\Omega
\eeq
where the metric function is
\beq
F(r) = 1-\frac{2M}{\Mpl^2 r} \left(1 + \frac{1}{2}\frac{\Mpl^2}{M^2}\right)
\label{gupmetric}
\eeq
Replacing $G = \Mpl^{-2}$, the horizon of the corresponding black hole is
\beq
R_H = 2GM + \frac{\beta}{M}
\eeq 
This encapsulates both the macroscopic (first term) and expected quantum behaviour (second term) of the horizon.   Indeed, a horizon of the form $R_H \sim \frac{1}{M}$ is that of a $(1+1)$-D black hole, which indicates the GUP-corrected black hole yields an effective dimensional reduction for Planck scale physics \cite{jrm1}.  This result is consistent with a variety of other gravitational theories, which suggests dimensional reduction might be a feature of a final quantum gravity model.

While the literature is replete with applications of the GUP to quantum gravity phenomenology, very few studies have considered large scale EUP effects on black holes (see {\it e.g.} \cite{bambi,eup1,eup2,Cadoni:2018dnd} and references therein).  Given a sufficiently large scale $\Ls$, it is possible that EUP modifications could be introduced to macroscopic black hole characteristics, including the horizon size and relativistic orbit parameters.  The scale $\Ls$ should be large enough so as not to impact solar system scale gravitation, but over larger ranges it will begin to have a stronger influence.  

In the spirit of the above discussion, one can thus propose an EUP-inspired metric that will account for new gravitational physics at some length scale $\Ls \gg \ellp$.  Following the corpuscular picture discussed earlier, consider a collection of $N$ gravitons, each of momentum uncertainty $\Delta p_g$, confined to a horizon length $\Delta x \sim R_S$. 
According to the EUP, 
\beq
\Delta p_g \sim \frac{\hbar}{R_S} \left(1 + \frac{\alpha R_S^2}{\Ls^2}\right)
\eeq
Noting that $N \frac{\hbar}{R_S} \sim \Mbh$, 
this expression becomes
\beq
\Delta P \sim M\left(1 +\frac{\alpha R_S^2}{\Ls^2}\right)
\eeq 
For small black holes with $R_S \ll \Ls$, $\Delta P$ is simply the (ADM) mass, and so it can be inferred that the above expression represents a total EUP-corrected mass,
\beq
{\cal M} = M\left(1 +\frac{4\alpha G^2M^2}{\Ls^2}\right)
\label{eupmass}
\eeq
assuming there is some EUP-based correction to the stress energy tensor, {\it i.e.}
\beq
{\cal M} = \int d^3x \sqrt{g}\left(T_{\rm GR}^{00} + T_{\rm EUP}^{00}\right)
\eeq
Replacing the mass term in the Schwarzschild metric with (\ref{eupmass}) thus yields
\beq
f(r) = 1-\frac{2GM}{r}\left(1 + \frac{4\alpha G^2M^2}{\Ls^2}\right)
\label{eupmetric}
\eeq
The horizon radius for (\ref{eupmetric}) is
\beq
R_H = 2GM+\frac{8\alpha G^3M^3}{\Ls^2}
\label{fullhorizon}
\eeq
which shows the large scale EUP correction becomes relevant when the horizon is of order $R_S \sim \Ls$.   If the EUP scale is commensurate with the Hubble length, $\Ls \sim L_H\sim 10^{26}~$m, the EUP corrections will not be observable for any known supermassive black hole.  

There is, however, no {\it a priori} reason that $\Ls$ should be so large.  In fact, observations clearly show deviations from GR at much smaller scales, particularly those attributed to dark matter.  In this case, $\Ls$ might well be of (sub)galactic scales, suggesting that dark matter effects may be somehow related to the EUP.  This is not dissimilar to ``dark forces'' induced by the Bose-Einstein condensate corpuscular graviton formalism, which has been shown to reproduce MOND-like accelerations at similar scales  \cite{Cadoni:2018dnd}. 

 \begin{center}
\begin{figure}[h]
\includegraphics[scale=0.35]{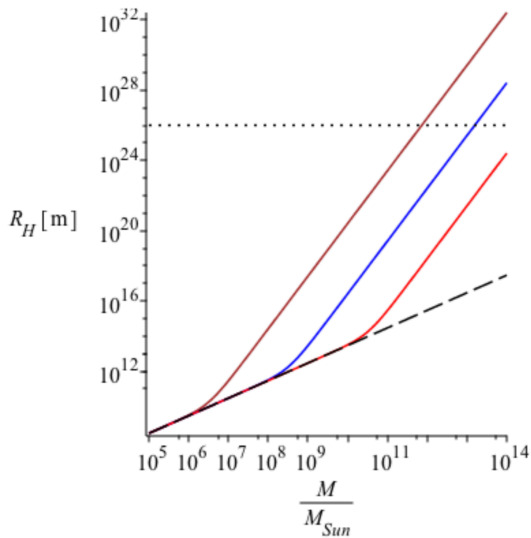}
\caption{EUP-modified horizon radius for black holes of mass $M$ and $\alpha =1$, with $\Ls = 10^{10}~$m (brown), $\Ls = 10^{12}~$m (blue), and $\Ls = 10^{14}~$m (red).  The standard Schwarzschild radius $R_S = 2GM/c^2$ is included (black dashed line).  The dotted line represents the Hubble scale, which provides an upper limit for horizon size. }
\label{fig1}
\end{figure}
\end{center}

A choice of $\Ls \sim 10^{12}-10^{14}~$m for the fundamental scale places the EUP enhancement near the upper range of most known supermassive black holes, whose masses have been estimated to be between $M\sim 10^6-10^{10}~\SM$.  Figure~\ref{fig1} show the resulting departures from the standard Schwarzschild radius for such $\Ls$ values.   The resulting deviations from General Relativity are potentially measurable by the EHT or future higher precision observation experiments.  As $\Ls$ is increased, the EUP effects emerge for SMBHs beyond the observed mass range.

For  Sgr A*, whose mass is approximately $M_{\rm SgrA*} \sim 10^6~M_\odot$ \cite{Johannsen:2015mdd}, a scale of $\Ls \sim 10^{10}~$m produces a horizon that is approximately 2.5 times as large as that predicted by GR.   Note that although this is commensurate with solar system scales, it does not impact the associated tests of GR such as the advance of perihelion of Mercury and the deflection of starlight since the EUP correction is $\frac{R_{S,\odot}^2}{\Ls^2} \sim 10^{-14}$. Unfortunately, this would correct the horizon of a $10^{10}$ solar mass SMBH to $R_H \approx 10^{21}~$m, which is ruled out by observation. 

An upper limit may be set by referring to the observational data for SMBHs to fix an upper limit on their possible horizon size.  The closest known object to Sgr A* is the star M2, whose highly elliptical orbit brings it within $1400R_{\rm S}$ \cite{GRAVcollab}, however structures have been observed as close as $3R_S$ from the centre \cite{Lu:2018uiv}.  For higher mass candidates, the constraints are somewhat looser.   The limb-brightened jet from the SMBH in M87 ($M =6.1\times 10^9~M_\odot$) has been measured within a sensitivity of $7~R_{\rm S,M87}$ \cite{Kim:2018hul}.  Additionally, the active region around 3C84 ($M = 2\times 10^9~M_\odot$) has been resolved to within approximately $300~R_{\rm S, 3C84}$ from the core \cite{Giovannini:2018abd}.  

This data implies that a lower bound of $\Ls \sim 10^{12}-10^{13}~$m can still potentially fit current observations.  For $\Ls \sim 10^{12}~$m, the horizon radius of Sgr A* would be bigger than the GR prediction by $0.01\%$, or about $1000~$km.  A value of $\Ls\sim 10^{13}~$m lowers to $10^{-4}\%$.  These differences are not resolvable in current observation.   Note that the corresponding horizon radius of a $10^{10}~M_\odot$ SMBH would be larger than the GR prediction by a factor of $10^3$ for $\Ls\sim10^{12}$~m, and a factor of 10 for $\Ls\sim 10^{13}~$m.  

This EUP metric imposes a maximum mass for a black hole in our universe, since the corresponding horizon cannot exceed the Hubble length $L_H \sim 10^{26}~$m.  For those SMBHs whose horizons exceed $\Ls$, the EUP term dominates and so one can estimate
\beq
R_H \sim L_H~~~\Longrightarrow~~~M_{\rm max} \sim \frac{(L_H \Ls^2)^{\frac{1}{3}}}{2 G}
\eeq
whose numerical values are found in Table~\ref{tab1}.  These values are many orders of magnitude above the predicted masses of known (putative) SMBHs, and thus this can be considered a self-check of the framework.  Of course, the upper bound on the size of possible black holes is likely several (many) orders of magnitude below the Hubble length, and can be further constrained by observation.

\begin{table}[h]
\begin{tabular}{lr}
$\Ls$~(m) ~~&~~ $M_{\rm max} ~(M_\odot)$ \\ \hline
$10^{11}$ & $3.4\times 10^{12}$\\
$10^{12}$ & $1.6\times 10^{13}$\\
$10^{13}$& $7.3\times 10^{13}$ \\
$10^{14}$&$3.4\times 10^{14}$ \\
\end{tabular}
\caption{Maximum possible EUP black hole mass as a function of fundamental scale factor $\Ls$.  Observational data rules out $\Ls < 10^{12}~$m.  }
\label{tab1}
\end{table}

Further justification for the choice of length scale can be elucidated by the Newtonian potential, derived from the weak field limit
\beq
g_{tt} = f(r) \approx 1+2\Phi(M,r)
\eeq
which in this case gives
\beq
\Phi(M,r) = \Phi_N(M,r)\left(1+\frac{R_S(M)^2}{\Ls^2}\right)
\label{potential}
\eeq
with $\Phi_N(M,r)$ the Newtonian potential for a mass $M$.  For the Milky Way galaxy, dark matter effects ({\it e.g.} the flattening of rotation curves) set on the order $r \sim 1~$kpc$= 10^{19}~$m.  The amount of mass interior to this radius is about $M\sim 10^{10} M_\odot$ \cite{yoshiaki}, for which the corresponding Schwarzschild radius is $R_S\sim 10^{13}~$m.  This is precisely the scale range of $\Ls$ considered herein, which would produce observable effects in SMBHs of that mass.

To close this section, it is appropriate to comment on the choice of sign for the EUP parameter.
If $\alpha = -1$, the corrected horizon will shrink relative to its general relativistic value,
\beq
R_H = 2GM-\frac{8\alpha G^3M^3}{\Ls^2}~~.
\eeq
This has drastic implications for SMBHs, since it implies more massive black holes have smaller horizons, leading to $R_H = 0$ for the maximum mass
\beq
M_{\rm max} = \frac{\Ls}{GM}
\eeq
Similarly, the weak field limit discussed above would result in a repulsive potential (\ref{potential}) for sufficiently large mass distributions, which is clearly ruled out by observation for the choice of $\Ls$ considered herein.

\section{Relativistic Orbits Around EUP Black Holes}
\begin{center}
\begin{figure}[h]
\includegraphics[scale=0.35,angle=-90]{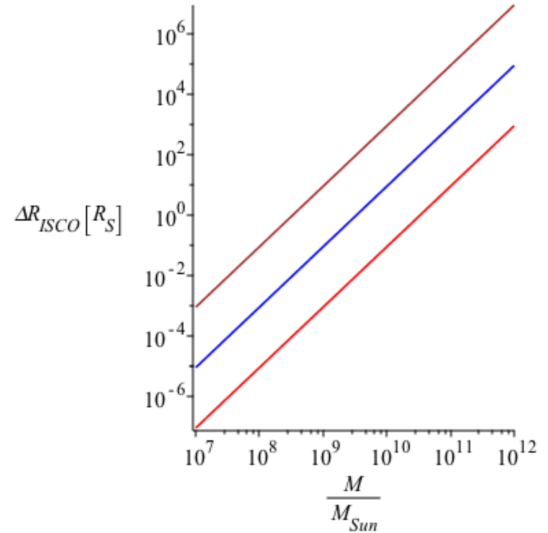}
\caption{The difference in ISCO radii between the EUP and standard GR, $\Delta R_{\rm ISCO} = R_{ISCO} - R_{ISCO, GR}$  in units of the SMBH Schwarzschild radius.  $\Ls = 10^{12}~$m (brown), for $\Ls = 10^{13}~$m (blue), and $\Ls = 10^{14}~$m (red).  }
\label{iscofig}
\end{figure}
\end{center}

Following the standard prescription, the equations for matter orbits around EUP black holes are straightforward to obtain.  Setting $\alpha = 1$, the effective potential is
\beq
V_{\rm matter} = -\frac{R_S}{2r} \left(1 + \frac{R_S^2}{\Ls^2}\right) +\frac{l^2}{2r^2}
- \frac{R_Sl^2}{2r^3}\left(1 + \frac{R_S^2}{\Ls^2}\right)
\label{matterveff}
\eeq
where $l$ is the matter particle's angular momentum.  Expressed in terms of the Schwarzschild radius, the stable orbits occur at
\beq
R_{\pm} = \frac{l^2}{R_S\left(1+\frac{2R_S^2}{\Ls^2}\right)}\left(1\pm\sqrt{1-\frac{3R_S^2\left(1+\frac{2R_S^2}{\Ls^2}\right)^2}{l^2}}\right)
\eeq
The critical angular momentum value for stable orbits is
\beq
l_{\rm ISCO} = \sqrt{3} R_S \left(1+\frac{R_S^2}{\Ls^2}\right)
\eeq
which gives the ISCO as
\beq
R_{\rm ISCO} = 3R_S+\frac{3R_S^3}{\Ls^2}
\eeq
Again, the EUP correction becomes relevant when $R_S \sim \Ls$.  As Figure~\ref{iscofig} indicates, the difference can be significant when $\Ls = 10^{12}~$m for SMBHs of masses greater than $M \sim 10^8~M_\odot$, while any significant different in the ISCOs will be apparent for SMBHs below $M=10^{12}~M_\odot$ for larger $\Ls$.

Similarly, the effective potential for photon orbits is obtained as
\beq
W_{\gamma} = \frac{1}{r^2}\left[1-\frac{R_S}{r}\left(1+\frac{R_S^2}{\Ls^2}\right)\right]
\eeq
and the photosphere is (as a function of $R_S$)
\beq
R^{\rm EUP}_{\rm ps} = 1.5 R_S + \frac{1.5 R_S^3}{\Ls^2}
\eeq
Note that the difference between EUP and GR photospheres is $\Delta R_{\rm ps} = \frac{1.5 R_S^3}{\Ls^2}$, which is different than that for the matter ISCOs.  This suggests a unique experimental test of the EUP-modified gravity considered herein, since the ISCO determines the accretion disk limits and the photosphere dictates the size of the black hole shadow.  

As a final note, the Hawking temperature of the EUP black hole can be readily computed as
\beq
T \sim f^\prime(R_H) \sim \frac{1}{8\pi G {\cal M}}
\eeq
In the limit $GM\ll \Ls$, the standard GR form is recovered, {\it i.e.}
\beq
T \sim \frac{1}{8\pi GM}
\eeq
but when $GM \gg \Ls$, one finds
\beq
T \sim \frac{\Ls^2}{32\pi G^3 M^3}
\eeq
Consequently, very large black holes have a much smaller temperature than that predicted by GR.  The crossover temperature occurs when
\beq
M_{\rm crit} \sim \frac{\Ls}{2G}
\eeq
Note this can also be written in terms of the Planck length as
\beq
M_{\rm crit} \sim \frac{\Ls}{\ellp^2}
\eeq
and thus is dependent on both of the two fundamental scales.

\section{Conclusions}
This paper has introduced a large mass scale correction to the Schwarzschild metric inspired by the Extended Uncertainty Principle.  All measurable properties of a black hole -- horizon, ISCO, and photosphere -- are modified by the addition of a term proportional  to $\frac{G^3M^3}{\Ls^2}$.  If $\Ls$ is on the order of $10^{12}-10^{14}~$m, such deviations will become relevant for supermassive black holes in the range of $M\sim 10^9-10^{11}~M_\odot$.  Although the deviations from GR will be much smaller for Sgr-A*, they could be measurable in the future with increased resolution.  

Future investigations will consider EUP modifications to Kerr black hole metrics, since the SMBHs in the galactic centres are significantly rotating.  This work is currently in progress.  Additional applications include corrections to weak-field (Newtonian) potentials, which can be used to model galactic rotation disks and determine if EUP gravity could partially mimic dark matter effects.  On its own, the $\frac{1}{r^2}$ form of the force will still produce a Keplerian velocity profile and not a flattening effect, so it does not completely negate the need for dark matter.  Since the Newtonian gravitational force from baryonic matter increases when the Schwarzschild radius of the enclosed mass exceeds $\Ls$, though, it will at least decrease the amount of dark matter needed.  This will be relevant to galactic and extra-galactic dynamics, particularly for objects outside the solar orbital radius.  

It is also of interest to determine how the modifications might look if $\Ls$ is near the value considered in this work, or even closer to the Hubble scale.  This will have implications for in the formation and dynamics of large scale structures, as well as associated cosmological models. 

Lastly, for completeness it is noted that the most general form of the uncertainty principle can be expressed as \cite{mannkempf}
\beq
\Delta x \Delta p \geq 1 + \beta \ellp^2 \Delta p^2 +\alpha \frac{\Delta x^2}{\Ls^2}~~.
\label{geup}
\eeq
which is often referred to as the Generalized Extended Uncertainty Principle (GEUP, \cite{eup2}).
This encompasses both quantum (GUP) and large (EUP) scale corrections to gravity, and thus provides ``bookends'' for quantum gravity effects (a natural choice for $\Ls$ is the Hubble scale, providing a ``dual'' counterpart to the Planck length $\ellp$).   If GUP effects are included as in Equation~(\ref{geup}), the corresponding black hole horizon would be
\beq
R_H = 2GM+\frac{8\alpha G^3M^3}{\Ls^2} + \frac{\beta}{M}
\eeq
which reduces to (\ref{fullhorizon}) in the large mass limit.   Since GUP effects will only be relevant near the Planck scale this form was not considered in the present paper, however future investigations can explore the possible new connections between quantum and classical gravitaiton.

\begin{acknowledgments}
I thank Roberto Casadio (University of Bologna) and Silke Britzen (MPIfR, Bonn) for insightful discussions.
\end{acknowledgments}


\begin{thebibliography}{99}

\bibitem{ligo} The LIGO Scientific Collaboration, {\tt https://www.ligo.org/}
\bibitem{eht} The Event Horizon Telescope collaboration, {\tt http://www.eventhorizontelescope.org/}

\bibitem{giddings1} S. Giddings, Phys.~Rev.~{\bf D 88}, 024018 (2013).
\bibitem{giddings2} S. Giddings, Phys.~Rev.~{\bf D 88}, 064023 (2013).
\bibitem{giddings3}S. Giddings and Y. Shi, Phys.~Rev.~{\bf D 89}, 124032 (2014).
\bibitem{giddings4} S. Giddings, Phys.~Rev.~{\bf D 90}, 124033 (2014).

\bibitem{bosonstar} D.~F.~Torres, S.~Capozziello and G.~Lambiase,
  %``A Supermassive scalar star at the galactic center?,''
  Phys.\ Rev.\ D {\bf 62}, 104012 (2000)
  [astro-ph/0004064].
\bibitem{planckstar}  
  C.~Rovelli and F.~Vidotto,
  %``Planck stars,''
  Int.\ J.\ Mod.\ Phys.\ D {\bf 23}, no. 12, 1442026 (2014)
  %doi:10.1142/S0218271814420267
  [arXiv:1401.6562 [gr-qc]].

\bibitem{fuzz1}  O.~Lunin and S.~D.~Mathur,
  %``AdS / CFT duality and the black hole information paradox,''
  Nucl.\ Phys.\ B {\bf 623}, 342 (2002)
 % doi:10.1016/S0550-3213(01)00620-4
  [hep-th/0109154].

\bibitem{fuzz2}  S.~D.~Mathur,
  %``The Fuzzball proposal for black holes: An Elementary review,''
  Fortsch.\ Phys.\  {\bf 53}, 793 (2005)
  %doi:10.1002/prop.200410203
  [hep-th/0502050].

\bibitem{fuzz3}
B.~Guo, S.~Hampton and S.~D.~Mathur,
  %``Can we observe fuzzballs or firewalls?,''
  JHEP {\bf 1807}, 162 (2018)
  %doi:10.1007/JHEP07(2018)162
  [arXiv:1711.01617 [hep-th]].
  
  
  


\bibitem{Cardoso:2017cqb} 
  V.~Cardoso and P.~Pani,
  %``Tests for the existence of black holes through gravitational wave echoes,''
  Nat.\ Astron.\  {\bf 1}, no. 9, 586 (2017)
  %doi:10.1038/s41550-017-0225-y
  [arXiv:1709.01525 [gr-qc]].
\bibitem{niayesh1} 
  J.~Abedi, H.~Dykaar and N.~Afshordi,
  Phys.\ Rev.\ D {\bf 96}, no. 8, 082004 (2017)
  [arXiv:1612.00266 [gr-qc]].

\bibitem{niayesh2} 
  Q.~Wang and N.~Afshordi,
  Phys.\ Rev.\ D {\bf 97}, no. 12, 124044 (2018)
  [arXiv:1803.02845 [gr-qc]].

\bibitem{niayesh3} 
  N.~Oshita and N.~Afshordi,
  arXiv:1807.10287 [gr-qc].

\bibitem{moffat2} 
  M.~A.~Green and J.~W.~Moffat,
  %``Extraction of black hole coalescence waveforms from noisy data,''
  doi:10.1016/j.physletb.2018.08.009
  arXiv:1711.00347 [astro-ph.IM].

\bibitem{DvaliGomez} 
G.~Dvali and C.~Gomez,
``Black Hole's Information Group'',
arXiv:1307.7630;
%``Black Holes as Critical Point of Quantum Phase Transition,''
Eur.\ Phys.\ J.\ C {\bf 74}, 2752 (2014)
[arXiv:1207.4059 [hep-th]];
%``Black Hole's 1/N Hair,''
Phys.\ Lett.\ B {\bf 719}, 419 (2013)
[arXiv:1203.6575 [hep-th]];
%``Landau-Ginzburg Limit of Black Hole's Quantum Portrait: Self Similarity and Critical Exponent,''
Phys.\ Lett.\ B {\bf 716}, 240 (2012)
[arXiv:1203.3372 [hep-th]];
%``Black Hole's Quantum N-Portrait,''
Fortsch.\ Phys.\  {\bf 61}, 742 (2013);
%[arXiv:1112.3359 [hep-th]];
G.~Dvali, C.~Gomez and S.~Mukhanov,
``Black Hole Masses are Quantized,''
arXiv:1106.5894 [hep-ph].
%


\bibitem{shadow1} L~.Amarilla, E.~F.~Eiroa, Phys.~Rev.~{\bf D 85}:064019 (2012).
%20%21
\bibitem{shadow3} A~.Grenzebach, V~.Perlick, C~.L\"ammerzahl, Phys.~Rev.~{\bf D 89}, 124004 (2014).
%22
\bibitem{shadow4} U.~Papnoi, F.~Atamurotov, S.~Ghosh, B.~Ahmedov, Phys.~Rev.~{\bf D 90}, 024073 (2014).
%23
\bibitem{shadow5}  Zi.~ Li, C.~Bambi, JCAP P {\bf 1401}:041 (2014).
%24
%\bibitem{shadow5a}  R.~S.~Lu {\it et al.}, Astrophys.\ J.\  {\bf 788}, 120 (2014).

\bibitem{shadow6} S.~Abdolrahimi, R.~Mann, C.~Tzounis, Phys.~Rev.~{\bf D 91}, 084052 (2015).
%25
\bibitem{moffat} J. W. Moffat, Eur.~Phys.~J.~{\bf C  75}:130 (2015).
%26
\bibitem{shadow8} P.~Cunha, C.~Herdeiro, E.~Radu, H.~Runarsson, Phys.\ Rev.\ Lett.\ {\bf 115}, 211102 (2015).

\bibitem{shadow9}   T.~Johannsen {\it et al.},  Phys.\ Rev.\ Lett.\  {\bf 116}, no. 3, 031101 (2016).

\bibitem{shadow10} 
  J.~R.~Mureika and G.~U.~Varieschi,
  %``Black hole shadows in fourth-order conformal Weyl gravity,''
  Can.\ J.\ Phys.\  {\bf 95}, no. 12, 1299 (2017)
%  doi:10.1139/cjp-2017-0241
  [arXiv:1611.00399 [gr-qc]].

\bibitem{mannkempf} 
  A.~Kempf, G.~Mangano and R.~B.~Mann,
  %``Hilbert space representation of the minimal length uncertainty relation,''
  Phys.\ Rev.\ D {\bf 52}, 1108 (1995)
 % doi:10.1103/PhysRevD.52.1108
  [hep-th/9412167].
\bibitem{Adler_1} R. J. Adler and D. I. Santiago, %On gravity and the Uncertainty Principle, 
Mod. Phys. Lett. {\bf A14}, 1371 (1999).

\bibitem{veneziano_1}
G. Veneziano, Europhys. Lett. {\bf 2}, 199 (1986). 
%4
\bibitem{veneziano_2} 
E. Witten, Phys.~Today {\bf 49N4}, 24-30 (1996).
\bibitem{veneziano_3}
F. Scardigli, Phys.~Lett.~ {\bf B452}, 39 (1999).
\bibitem{veneziano_4}
D.  J. Gross and P. F. Mende, Nuc. Phys. {\bf B303}, 407 (1988).
\bibitem{veneziano_5}
D. Amati, M. Ciafaloni and G. Veneziano, Phys, Lett. {\bf B216}, 41 (1989).
\bibitem{veneziano_6}
 T. Yoneya, Mod. Phys. Lett. {\bf A4}, 1587 (1989)

\bibitem{ashtekar_1}
A. Ashtekar, S. Fiarhurst and J. L. Willis, Class. Quant. Grav. {\bf 20}, 1031 (2003).
%G. M. Hossain, V. Husain and S. S. Seahra, Class. Quant. Grav. {\bf 207}, 165013 (2010).
%3
\bibitem{ashtekar_2}
%A. Ashtekar, S. Fiarhurst and J. L. Willis, Class. Quant. Grav. {\bf 20}, 1031 (2003);
G. M. Hossain, V. Husain and S. S. Seahra, Class. Quant. Grav. {\bf 27}, 165013 (2010).

\bibitem{nicolini} M. Isi, J. Mureika and P. Nicolini, JHEP~{\bf 1311}, 139 (2013).

\bibitem{majid} 
S. Majid, 
%Scaling limit of the non-commutative black hole, 
J. Phys. Conf. Ser. 284, 012003 (2011).

%5
\bibitem{maggiore_1}
M. Maggiore, Phys. Lett. {\bf B 304}, 65 (1993).
\bibitem{maggiore_2}
M. Maggiore, Phys. Lett. {\bf B 319}, 83 (1993).
\bibitem{maggiore_3}
M. Maggiore,  Phys. Rev. {\bf D 49}, 5182 (1994).

\bibitem{Adler_2}R. J. Adler, P. Chen and D. I. Santiago, %The Generalized Uncertainty Principle and black hole remnants, 
Gen. Rel. Grav. {\bf 33}, 2101 (2001).
\bibitem{Adler_3} P. Chen and R. J. Adler, %Black hole remnants and dark matter 
Nucl.Phys.Proc.Suppl. \textbf{124}  103 (2003). %P Chen, Might dark matter be actually black? arXiv:astro-ph/0303349;
%\bibitem{Adler_4}R. J. Adler, %Six easy roots to the Planck scale 
%Am.\ J.\ Phys.\  {\bf 78}, 925 (2010).

\bibitem{gup1} 
  A.~N.~Tawfik and A.~M.~Diab,
  %``Generalized Uncertainty Principle: Approaches and Applications,''
  Int.\ J.\ Mod.\ Phys.\ D {\bf 23}, no. 12, 1430025 (2014)
  doi:10.1142/S0218271814300250
  [arXiv:1410.0206 [gr-qc]].
\bibitem{gup2} B.~J.~Carr,
  %``The Black Hole Uncertainty Principle Correspondence,''
  Springer Proc.\ Phys.\  {\bf 170}, 159 (2016)
  %doi:10.1007/978-3-319-20046-0_19
  [arXiv:1402.1427 [gr-qc]].
\bibitem{gup3} B.~J.~Carr, Mod.\ Phys.\ Lett.\ {\bf A28} (2013) 1340011


  
\bibitem{mijmpn}
  M.~Isi, J.~Mureika and P.~Nicolini,
  %``Self-Completeness and the Generalized Uncertainty Principle,''
  JHEP {\bf 1311}, 139 (2013)
  doi:10.1007/JHEP11(2013)139
  [arXiv:1310.8153 [hep-th]].
  
  \bibitem{gupextrad} 
  S.~Köppel, M.~Knipfer, M.~Isi, J.~Mureika and P.~Nicolini,
  %``Generalized uncertainty principle and extra dimensions,''
  arXiv:1703.05222 [hep-th].

\bibitem{bcjmpn} 
  B.~J.~Carr, J.~Mureika and P.~Nicolini,
  %``Sub-Planckian black holes and the Generalized Uncertainty Principle,''
  JHEP {\bf 1507}, 052 (2015)
 % doi:10.1007/JHEP07(2015)052
  [arXiv:1504.07637 [gr-qc]].
  %%CITATION = doi:10.1007/JHEP07(2015)052;%%
  %33 citations counted in INSPIRE as of 01 Aug 2018
  
  \bibitem{gup4} 
  F.~Scardigli, M.~Blasone, G.~Luciano and R.~Casadio,
  %``Modified Unruh effect from Generalized Uncertainty Principle,''
  arXiv:1804.05282 [hep-th].
  
\bibitem{jrm1} J.~R.~Mureika, Phys.~Lett.~{\bf B 716}, 171-175 (2012). [arXiv:1204.3619 [gr-qc]].
%17


\bibitem{bambi} 
  C.~Bambi and F.~R.~Urban,
  Class.\ Quant.\ Grav.\  {\bf 25}, 095006 (2008)
  [arXiv:0709.1965 [gr-qc]].

\bibitem{eup1}  S.~Mignemi,
  %``Extended uncertainty principle and the geometry of (anti)-de Sitter space,''
  Mod.\ Phys.\ Lett.\ A {\bf 25}, 1697 (2010)
  % doi:10.1142/S0217732310033426
  [arXiv:0909.1202 [gr-qc]].

\bibitem{eup2}  T.~Zhu, J.~R.~Ren and M.~F.~Li,
  %``Influence of Generalized and Extended Uncertainty Principle on Thermodynamics of FRW universe,''
  Phys.\ Lett.\ B {\bf 674}, 204 (2009)
 % doi:10.1016/j.physletb.2009.03.020
  [arXiv:0811.0212 [hep-th]].

\bibitem{Cadoni:2018dnd} 
  M.~Cadoni, R.~Casadio, A.~Giusti and M.~Tuveri,
  Phys.\ Rev.\ D {\bf 97}, no. 4, 044047 (2018)
  [arXiv:1801.10374 [gr-qc]].

\bibitem{Johannsen:2015mdd} 
  T.~Johannsen,
  %``Sgr A* and General Relativity,''
  Class.\ Quant.\ Grav.\  {\bf 33}, no. 11, 113001 (2016)
 % doi:10.1088/0264-9381/33/11/113001
  [arXiv:1512.03818 [astro-ph.GA]].

\bibitem{GRAVcollab}
  R.~Abuter {\it et al.} [GRAVITY Collaboration],
  %``Detection of the gravitational redshift in the orbit of the star S2 near the Galactic centre massive black hole,''
  arXiv:1807.09409 [astro-ph.GA].

\bibitem{Lu:2018uiv} 
  R.~S.~Lu {\it et al.},
  %``Detection of intrinsic source structure at ~3 Schwarzschild radii with Millimeter-VLBI observations of SAGITTARIUS A*,''
  Astrophys.\ J.\  {\bf 859}, no. 1, 60 (2018)
  %doi:10.3847/1538-4357/aabe2e
  [arXiv:1805.09223 [astro-ph.GA]].

\bibitem{Kim:2018hul} 
  J.-Y.~Kim {\it et al.},
  %``The limb-brightened jet of M87 down to 7 Schwarzschild radii scale,''
  arXiv:1805.02478 [astro-ph.GA].
  
\bibitem{Giovannini:2018abd} 
  G.~Giovannini {\it et al.},
  %``A wide and collimated radio jet in 3C 84 on the scale of a few hundred gravitational radii,''
  Nat.\ Astron.\  {\bf 2}, no. 6, 472 (2018)
 % doi:10.1038/s41550-018-0431-2
  [arXiv:1804.02198 [astro-ph.GA]].

\bibitem{yoshiaki} Y.~Sofue, ``Mass Distribution and Rotation Curve in the Galaxy'',  In {\it Planets, Stars and Stellar Systems}, T.~Oswalt, G.~Gilmore (eds.), Springer, Dordrecht (2013).
\end{thebibliography}
\end{document}